\documentclass[twocolumn, conference, 10pt]{IEEEtran}
\usepackage{amsmath,amssymb,amsthm}
\usepackage{bm}
\usepackage{url}

\newcommand{\Tr}{\mathrm{Tr}}

\theoremstyle{definition}

\theoremstyle{remark}

\title{Holographic Transformation for Quantum Factor Graphs}
\author{\IEEEauthorblockN{Ryuhei~Mori}
\IEEEauthorblockA{Department of Mathematical and Computing Science,\\
Tokyo Institute of Technology, Meguro-ku Ookayama, Tokyo, Japan 152-8552\\
email: mori@is.titech.ac.jp}}

\begin{document}
\maketitle
\begin{abstract}
Recently, a general tool called a holographic transformation, which transforms an expression of the partition function to another form,
has been used for polynomial-time algorithms and for improvement and understanding of the belief propagation.
In this work, the holographic transformation is generalized to quantum factor graphs.
\end{abstract}

\section{Introduction}
The computation of the partition function is one of the most important problem in statistical physics, machine learning,
computer science and information theory~\cite{mezard2009ipa}.
Recently, Valiant invented the holographic transformation for transforming the expression of the partition function
in order to obtain polynomial-time algorithms for planar graphs~\cite{valiant2008holographic}.
The idea of the holographic transformation explains many well-known identities, e.g.,
high-temperature expansion, MacWilliams identity, loop calculus, etc.~\cite{forney2011partition}, \cite{chernyak2007loop}, \cite{mori2013lcn}.
In this work, the holographic transformation is generalized to quantum factor graphs, which is restricted quantum graphical model
suggested in~\cite{leifer2008quantum}.
Although problems from quantum statistical physics are not included in our setting,
a decoding problems of quantum error correcting codes can be represented by a quantum factor graph~\cite{leifer2008quantum}.
Hence, the quantum generalization in this paper may be considered as the first step for generalizing loop calculus for quantum error correcting codes.
This paper suggests the generalization of the holographic transformation to quantum factor graphs,
but does not include any particular example.

\section{Factor graphs}
A factor graph is a bipartite graph defining a probability measure.
A factor graph consists of variable nodes, factor nodes and edges between a variable node and a factor node.
Let $V$ be the set of variable nodes and $F$ be the set of factor nodes.
Let $E\subseteq V\times F$ be the set of edges.
For a variable node $i\in V$, $\partial i\subseteq F$ denotes the set of neighborhoods of $i$.
In the same way $\partial a\subseteq V$ is defined for $a\in F$.
For each variable node $i\in V$, there is an associated finite set $\mathcal{X}_i$ and an associated function $f_i\colon \mathcal{X}_i\to \mathbb{R}_{\ge 0}$.
For each factor node $a\in F$, there is an associated function $f_a\colon\prod_{i\in\partial a}\mathcal{X}_i\to\mathbb{R}_{\ge 0}$.
Let $\bm{x}_{V'}\in\prod_{i\in V'}\mathcal{X}_i$ be variables corresponding to a subset $V'\subseteq V$ of variable nodes.
Then, the probability measure on $\mathcal{X}:=\prod_{i\in V}\mathcal{X}_i$ associated with the factor graph $G=(V,F,E,(f_i)_{i\in V}, (f_a)_{a\in F})$ is defined as
\begin{align*}
p(\bm{x}) &= \frac1{Z(G)} \prod_{a\in F} f_a(\bm{x}_{\partial a})\prod_{i\in V} f_i(x_i)\\
Z(G) &:= \sum_{\bm{x}\in\mathcal{X}} \prod_{a\in F} f_a(\bm{x}_{\partial a})\prod_{i\in V} f_i(x_i).
\end{align*}
Here, the constant $Z(G)$ for the normalization is called the partition function, which plays an important role in statistical physics,
machine learning, computer science and information theory~\cite{mezard2009ipa}.

\section{Holographic transformation for factor graphs}
In this section, we briefly review the holographic transformation for classical factor graphs.
Let $\mathcal{X}_{i,a}:=\mathcal{X}_i$ for all $(i,a)\in E$.
Let $\phi_{i,a}\colon \mathcal{X}_i\times\mathcal{X}_{i,a} \to \mathbb{R}$ and
$\hat{\phi}_{i,a}\colon \mathcal{X}_{i,a}\times\mathcal{X}_{i,a} \to \mathbb{R}$ be mappings for each $(i,a)\in E$ satisfying
\begin{equation}
\sum_{y\in\mathcal{X}_i} \phi_{i,a}(x, y)\hat{\phi}_{i,a}(y, z) = \delta(x,z)
\label{eq:invc}
\end{equation}
where $\delta(x,z)$ takes 1 if $x=z$ and 0 otherwise.
Let $\mathcal{Y}:=\prod_{(i,a)\in E}\mathcal{X}_{i,a}$
Then, it holds
\begin{align*}
Z(G) &= \sum_{\bm{x}\in\mathcal{X}} \prod_{a\in F} f_a(\bm{x}_{\partial a})\prod_{i\in V} f_i(x_i)\\
&= \sum_{\bm{x}\in\mathcal{X}, \bm{z}\in\mathcal{Y}} \prod_{a\in F} f_a(\bm{z}_{\partial a, a})\prod_{i\in V} f_i(x_i)
\prod_{(i,a)\in E}\delta(x_i, z_{i,a})\\
&= \sum_{\bm{x}\in\mathcal{X}, \bm{z}\in\mathcal{Y}, \bm{y}\in\mathcal{Y}} \prod_{a\in F} f_a(\bm{z}_{\partial a, a})\prod_{i\in V} f_i(x_i)\\
&\quad\cdot\prod_{(i,a)\in E}\phi_{i,a}(x_i, y_{i,a})\hat{\phi}_{i,a}(y_{i,a}, z_{i,a})\\
&= \sum_{\bm{y}\in\mathcal{Y}}
\prod_{a\in F} \left(\sum_{\bm{z}_{\partial a, a}\in\mathcal{Y}_{\partial a, a}} f_a(\bm{z}_{\partial a, a})\prod_{i\in\partial a}\hat{\phi}_{i,a}(y_{i,a}, z_{i,a})\right)\\
&\quad\cdot \prod_{i\in V} \left(\sum_{x_i\in\mathcal{X}_i}f_i(x_i)\prod_{a\in\partial i}\phi_{i,a}(x_i, y_{i,a})\right)
\end{align*}
where $\bm{z}_{\partial a, a}:=(z_{i,a})_{i\in\partial a}$.
By letting
\begin{align*}
\hat{f}_a(\bm{y}_{\partial a, a})&:=
\sum_{\bm{z}_{\partial a, a}\in\mathcal{Y}_{\partial a, a}} f_a(\bm{z}_{\partial a, a})
 \prod_{i\in\partial a}\hat{\phi}_{i,a}(y_{i,a}, z_{i,a})\\
\hat{f}_i(\bm{y}_{i,\partial i}) &:= \sum_{x_i\in\mathcal{X}_i}f_i(x_i)\prod_{a\in\partial i}\phi_{i,a}(x_i, y_{i,a})
\end{align*}
where $\bm{y}_{i,\partial i}:=(y_{i,a})_{a\in\partial i}$,
one obtains
\begin{equation}
Z(G) = \sum_{\bm{y}\in\mathcal{Y}} \prod_{a\in F} \hat{f}_a(\bm{y}_{\partial a, a})\prod_{i\in V} \hat{f}_i(\bm{y}_{i,\partial i}).
\label{eq:cholant}
\end{equation}
This equality is called the Holant theorem~\cite{valiant2008holographic}, \cite{5695119}, which explains many known equalities~\cite{forney2011partition}.

\section{Quantum factor graphs}
There are several quantum graphical models understood as generalizations of classical factor graphs.
In quantum physics, quantum state is expressed by a positive semidefinite trace-1 matrix, called a density matrix.
The conventional matrix product cannot be used for factor graph directly since a product of two positive semidefinite matrices
is not necessarily positive semidefinite.
By considering the conditional independence for quantum states~\cite{leifer2008quantum},
the most natural generalization $\odot$ of the products in the classical factor graph would be
\begin{align*}
(\Lambda\odot \Lambda')|\psi\rangle &:= 0,&\text{if } |\psi\rangle \notin S\\
(\Lambda\odot \Lambda')|\psi\rangle &:= \exp\{\log \Lambda_{|S} + \log \Lambda'_{|S}\}|\psi\rangle
,&\text{if } |\psi\rangle \in S
\end{align*}
where $S$ is the intersection of the supports of $\Lambda$ and $\Lambda'$,
and where $\Lambda_{|S}$ and $\Lambda'_{|S}$ are the restriction of $\Lambda$ and $\Lambda'$, respectively, to $S$.
Here, $\Lambda$ and $\Lambda'$ must be positive semidefinite.
Obviously $\Lambda\odot\Lambda'$ is also positive semidefinite.
While the product $\odot$ is commutative and associative,
the product $\odot$ is not distributive with the partial trace in general~\cite{leifer2008quantum}.
Hence, in this paper, we do not deal with the quantum graphical model using the product $\odot$ although it includes
problems from quantum statistical physics.

The Suzuki-Trotter approximation for $\odot$ gives a set of definitions of products $\star^{(n)}$ as
\begin{equation*}
\Lambda \star^{(n)} \Lambda'
:=
\left(\Lambda^{\frac1{2n}} \Lambda'^{\frac1{n}}\Lambda^{\frac1{2n}} \right)^n.
\end{equation*}
The product $\odot$ is obtained as the limit of $\star^{(n)}$
\begin{equation*}
\Lambda\odot\Lambda' = \lim_{n\to\infty} \Lambda\star^{(n)}\Lambda'.
\end{equation*}
While $\Lambda\star^{(n)}\Lambda'$ is positive semidefinite if $\Lambda$ and $\Lambda'$ are positive semidefinite,
$\star^{(n)}$ is neither commutative nor associative.
However, $\star := \star^{(1)}$ is useful for guaranteeing the distributive law with the partial trace, i.e.,
$\Tr(\Lambda_{AB}\star\Lambda_{BC})=\Tr_{\mathcal{H}_B}(\Tr_{\mathcal{H}_A}(\Lambda_{AB})\star\Tr_{\mathcal{H}_C}(\Lambda_{BC}))$
where $\Lambda_{AB}$ and $\Lambda_{BC}$ are positive semidefinite operators acting on Hilbert spaces $\mathcal{H}_A\otimes\mathcal{H}_B$
and $\mathcal{H}_B\otimes\mathcal{H}_C$, respectively.
Hence, in this paper, we deal with the quantum factor graph using the product $\star$ defined in~\cite{leifer2008quantum}.
Let $V$, $F$, $E$ be the sets of variable nodes, factor nodes and edges as with the classical factor graph.
For each variable node $i\in V$, there is an associated Hilbert space $\mathcal{H}_i$ and an associated positive semidefinite operator $f_i$ on $\mathcal{H}_i$.
For each factor node $a\in F$, there is an associated positive semidefinite operator $f_a$ on $\mathcal{H}_{\partial a}:=\bigotimes_{i\in\partial a}\mathcal{H}_i$.
Then, the density operator on $\mathcal{H}:=\bigotimes_{i\in V} \mathcal{H}_i$ associated with the quantum factor graph
$G=(V,F,E,(f_i)_{i\in V}, (f_a)_{a\in F})$ is defined as
\begin{equation*}
\begin{split}
\rho &= \frac1{Z(G)} \left(\prod_{a\in F} f_a\right)\star\left(\bigotimes_{i\in V} f_i\right)\\
Z(G) &:= \Tr_{\mathcal{H}}\left(\left(\prod_{a\in F} f_a\right)\star\left(\bigotimes_{i\in V} f_i\right)\right)\\
&= \Tr_{\mathcal{H}}\left(\left(\prod_{a\in F} f_a\right)\left(\bigotimes_{i\in V} f_i\right)\right)
\end{split}
\end{equation*}
where $(f_a)_{a\in F}$ are mutually commute with respect to the conventional matrix product.
In this paper, all Hilbert spaces are assumed to be finite dimensional.
This quantum model includes decoding problem of quantum error correcting codes~\cite{leifer2008quantum}.

\section{Non-commutative holographic transformation for quantum factor graphs}
In this section, the holographic transformation and the Holant theorem in \cite{valiant2008holographic}, \cite{5695119}, \cite{mori2013lcn} for classical factor graphs
is generalized to quantum factor graphs.
Let $\mathcal{H}'_{i,a}$ and $\hat{\mathcal{H}}_{i,a}$ be new Hilbert spaces
which are isomorphic to $\mathcal{H}_i$.
Let $q_i$ be the dimension of $\mathcal{H}_i$ and $(|e_j\rangle_{\mathcal{H}_i})_{j=1,\dotsc,q_i}$
be a basis for $\mathcal{H}_i$ for all $i\in V$.
Let $f'_a$ be the same as $f_a$ but acting on $\mathcal{H}'_{\partial a}:=\bigotimes_{i\in\partial a}\mathcal{H}'_{i,a}$ for all $a\in F$.
Let $\mathcal{B}(\mathcal{H}_A)$ be the set of linear operators acting on a Hilbert space $\mathcal{H}_A$.
The set $\mathcal{B}(\mathcal{H}_A)$ of linear operators can also be regarded as a linear space
with respect to the conventional summation and scalar multiplication.
Let $\Phi_{i,a}$ and $\hat{\Phi}_{i,a}$ be linear maps from $\mathcal{B}(\hat{\mathcal{H}}_{i,a})$ to $\mathcal{B}(\mathcal{H}_i)$
and from $\mathcal{B}(\mathcal{H}'_{i,a})$ to $\mathcal{B}(\hat{\mathcal{H}}_{i,a})$, respectively, for all $(i,a)\in E$.
For any linear map $T\colon \mathcal{B}(\mathcal{H}_A)\to \mathcal{B}(\mathcal{H}_B)$,
there is the Choi-Jamiolkowski representation $\tau\in\mathcal{B}(\mathcal{H}_B\otimes\mathcal{H}_A)$ of $T$, i.e.,
\begin{equation*}
\tau := \sum_{k,l} T(|e_k\rangle_{\mathcal{H}_A}\langle e_l|_{\mathcal{H}_A})\otimes|e_k\rangle_{\mathcal{H}_A}\langle e_l|_{\mathcal{H}_A}
\end{equation*}
which satisfies
\begin{equation*}
T(G) = \Tr_{\mathcal{H}_{A}}\left(\tau \left(I_{\mathcal{H}_B} \otimes G^T\right)\right)
\end{equation*}
for any $G\in\mathcal{B}(\mathcal{H}_A)$
where $I_{\mathcal{H}_B}$ denotes the identity operator on $\mathcal{H}_B$ and ${}^T$ denotes the transpose of linear map~\cite{wolf2012quantum}.
Let $\phi_{i,a}\in\mathcal{B}(\mathcal{H}_i\otimes\hat{\mathcal{H}}_{i,a})$ and
$\hat{\phi}_{i,a}\in\mathcal{B}(\hat{\mathcal{H}}_{i,a}\otimes\mathcal{H}'_{i,a})$
be the Choi-Jamiolkowski representations for $\Phi_{i,a}$ and $\hat{\Phi}_{i,a}$, respectively, for all $(i,a)\in E$.

We assume that $(\Phi_{i,a})_{a\in\partial i}$ are mutually commute with respect to the conventional matrix product for $i\in V$.\footnote{Although this condition may not be necessary, we assume it for the simplicity}
Furthermore, we assume that it holds
$\Phi_{i,a}\hat{\Phi}_{i,a} = \mathrm{id}_{\mathcal{H}_i,\mathcal{H}'_{i,a}}$
for all $(i,a)\in E$
where
\begin{equation*}
\mathrm{id}_{\mathcal{H}_i,\mathcal{H}'_{i,a}}\left(|e_k\rangle_{\mathcal{H}'_{i,a}}\langle e_l|_{\mathcal{H}'_{i,a}}\right)
:=|e_k\rangle_{\mathcal{H}_{i}}\langle e_l|_{\mathcal{H}_{i}}
\end{equation*}
for all $k,l \in\{1,2,\dotsc,q_i\}$.
It is easy to verify that
this condition is equivalent to
\begin{equation}
\Tr_{\hat{\mathcal{H}}_{i, a}}\left(\hat{\phi}^T_{i,a}\phi_{i,a}\right) =  \mathbb{F}_{i,a} :=
\sum_{k,l}
|e_k\rangle_{\mathcal{H}_i} \langle e_l|_{\mathcal{H}_i} 
\otimes
|e_l\rangle_{\mathcal{H}'_{i,a}} \langle e_k|_{\mathcal{H}'_{i,a}}.
\label{eq:inv0}
\end{equation}
Let $\hat{\mathcal{H}}:=\bigotimes_{(i,a)\in E}\hat{\mathcal{H}}_{i,a}$ and $\mathcal{H}':=\bigotimes_{(i,a)\in E}\mathcal{H}'_{i,a}$.
Then, one obtains
\begin{align*}
&Z(G)=
\Tr_{\mathcal{H}}
\left(\prod_{a\in F} \Tr_{\mathcal{H}'_{\partial a}}\left(f'_a\bigotimes_{i\in \partial a} \mathbb{F}_{i,a}\right) \bigotimes_{i\in V} f_i\right)\\
&=
\Tr_{\mathcal{H}\otimes\hat{\mathcal{H}}\otimes\mathcal{H}'}
\left(\prod_{a\in F} \left(f'_a\bigotimes_{i\in \partial a}\left(\hat{\phi}^T_{i,a}\phi_{i,a}\right)\right) \bigotimes_{i\in V} f_i\right)\\
&=
\Tr_{\hat{\mathcal{H}}}
\Biggl(\bigotimes_{a\in F} \Tr_{\mathcal{H}'_{\partial a}}\left(f'_a\left(\bigotimes_{i\in\partial a}\hat{\phi}^T_{i,a}\right)\right)\\
&\qquad\cdot
\bigotimes_{i\in V} \Tr_{\mathcal{H}_i}\left(\left(\prod_{a\in\partial i} \phi_{i,a}\right)f_i\right) \Biggr).
\end{align*}
For linear maps $T\colon \mathcal{B}(\mathcal{H}_A)\to\mathcal{B}(\mathcal{H}_B)$
and $T'\colon \mathcal{B}(\mathcal{H}_C)\to\mathcal{B}(\mathcal{H}_B)$,
$T\bar{\otimes} T'\colon \mathcal{B}(\mathcal{H}_A)\otimes\mathcal{B}(\mathcal{H}_C)\to\mathcal{B}(\mathcal{H}_B)$ is defined as
\begin{align*}
&(T \bar{\otimes} T')(|e_k\rangle_{\mathcal{H}_A}\langle e_l|_{\mathcal{H}_A}\otimes|e_{k'}\rangle_{\mathcal{H}_C}\langle e_{l'}|_{\mathcal{H}_C})\\
&\quad:=
T(|e_k\rangle_{\mathcal{H}_A}\langle e_l|_{\mathcal{H}_A})
T'(|e_{k'}\rangle_{\mathcal{H}_C}\langle e_{l'}|_{\mathcal{H}_C}).
\end{align*}
Let $T^*$ be the adjoint map of $T$, i.e., $\Tr(BT(A))=\Tr(T^*(B)A)$~\cite{wolf2012quantum}.
Then, one obtains
\begin{align}
Z(G) &= \Tr_{\hat{\mathcal{H}}}\left(\bigotimes_{a\in F}\hat{f}^T_a\bigotimes_{i\in V} \hat{f}^T_i\right)
= \Tr_{\hat{\mathcal{H}}}\left(\bigotimes_{a\in F}\hat{f}_a\bigotimes_{i\in V} \hat{f}_i\right)
\label{eq:holant}
\end{align}
where
\begin{align*}
\hat{f}_a &:= \Tr_{\mathcal{H}'_{\partial a}}\left( \left(\bigotimes_{i\in\partial a} \hat{\phi}_{i,a}\right)f'^T_a\right)
= \left(\bigotimes_{i\in\partial a}\hat{\Phi}_{i,a}\right)\left(f'_a\right)\\
\hat{f}_i &:= \Tr_{\mathcal{H}_i}\left( f^T_i\left(\prod_{a\in\partial i}\phi^T_{i,a}\right)\right)
= \left(\overline{\bigotimes_{a\in\partial i}}\Phi_{i,a}\right)^* \left(f_i\right).
\end{align*}
The equation~\eqref{eq:holant} can be regarded as a quantum generalization of the Holant theorem~\eqref{eq:cholant}.

Note that for the classical case, $(f_a)_{a\in F}$ and $(f_i)_{i\in V}$ are diagonal with respect to some basis
$\{\bigotimes_{i\in V}|e_{k_i}\rangle_{\mathcal{H}_i}\mid \forall i\in V,\, k_i=1,\dotsc,q_i\}$.
In this case, the condition~\eqref{eq:inv0} can be replaced by a different condition
\begin{equation*}
\left(\Phi_{i,a}\hat{\Phi}_{i,a}\right)\left(|e_k\rangle_{\mathcal{H}'_{i,a}}\langle e_l|_{\mathcal{H}'_{i,a}}\right)
:=\delta(k,l)|e_k\rangle_{\mathcal{H}_{i}}\langle e_l|_{\mathcal{H}_{i}}
\end{equation*}
for all $k,l \in\{1,2,\dotsc,q_i\}$
or equivalently,
\begin{equation*}
\Tr_{\hat{\mathcal{H}}_{i, a}}(\hat{\phi}^T_{i,a}\phi_{i,a}) = 
\sum_{j=1}^{q_i}|e_j\rangle_{\mathcal{H}_i} \langle e_j|_{\mathcal{H}_{i}} \otimes |e_j\rangle_{\mathcal{H}'_{i,a}}  \langle e_j|_{\mathcal{H}'_{i,a}}.
\end{equation*}
If $(\phi_{i,a})_{(i,a)\in E}$ and $(\hat{\phi}_{i,a})_{(i,a)\in E}$ are restricted to be diagonal with respect to the same basis,
the above condition corresponds to~\eqref{eq:invc}.

\section{Conclusion and future work}
The holographic transformation is generalized to quantum factor graphs.
Finding particular examples is the most important future work.
The quantum holographic transformation for classical factor graphs is also interesting framework.

\section*{Acknowledgment}
This work was supported by MEXT KAKENHI Grant Number 24106008.

\bibliographystyle{IEEEtran}
\bibliography{IEEEabrv,biblio}

\begin{thebibliography}{1}
\providecommand{\url}[1]{#1}
\csname url@samestyle\endcsname
\providecommand{\newblock}{\relax}
\providecommand{\bibinfo}[2]{#2}
\providecommand{\BIBentrySTDinterwordspacing}{\spaceskip=0pt\relax}
\providecommand{\BIBentryALTinterwordstretchfactor}{4}
\providecommand{\BIBentryALTinterwordspacing}{\spaceskip=\fontdimen2\font plus
\BIBentryALTinterwordstretchfactor\fontdimen3\font minus
  \fontdimen4\font\relax}
\providecommand{\BIBforeignlanguage}[2]{{%
\expandafter\ifx\csname l@#1\endcsname\relax
\typeout{** WARNING: IEEEtran.bst: No hyphenation pattern has been}%
\typeout{** loaded for the language `#1'. Using the pattern for}%
\typeout{** the default language instead.}%
\else
\language=\csname l@#1\endcsname
\fi
#2}}
\providecommand{\BIBdecl}{\relax}
\BIBdecl

\bibitem{mezard2009ipa}
M.~Mezard and A.~Montanari, \emph{{Information, Physics and
  Computation}}.\hskip 1em plus 0.5em minus 0.4em\relax Oxford University
  Press, 2009.

\bibitem{valiant2008holographic}
L.~G. Valiant, ``Holographic algorithms,'' \emph{SIAM Journal on Computing},
  vol.~37, no.~5, pp. 1565--1594, 2008.

\bibitem{forney2011partition}
G.~D. Forney, Jr. and P.~O. Vontobel, ``Partition functions of normal factor
  graphs,'' in \emph{Proc. 2011 IEEE Inf. Theory and App. Workshop, La Jolla,
  CA}, 2011.

\bibitem{chernyak2007loop}
V.~Y. Chernyak and M.~Chertkov, ``Loop calculus and belief propagation for
  $q$-ary alphabet: Loop tower,'' in \emph{Proc. 2007 IEEE Int. Symposium on
  Inform. Theory, Nice, France}, Jun. 24--29, 2007, pp. 316--320.

\bibitem{mori2013lcn}
\BIBentryALTinterwordspacing
R.~Mori, ``Loop calculus for non-binary alphabets using concepts of information
  geometry,'' 2013, submitted for the publication in {IEEE} Trans. Inf. Theory.
  [Online]. Available: \url{http://arxiv.org/abs/1309.6550}
\BIBentrySTDinterwordspacing

\bibitem{leifer2008quantum}
M.~Leifer and D.~Poulin, ``Quantum graphical models and belief propagation,''
  \emph{Annals of Physics}, vol. 323, no.~8, pp. 1899--1946, 2008.

\bibitem{5695119}
A.~Al-Bashabsheh and Y.~Mao, ``Normal factor graphs and holographic
  transformations,'' \emph{{IEEE} Trans. Inf. Theory}, vol.~57, no.~2, pp. 752
  --763, Feb. 2011.

\bibitem{wolf2012quantum}
\BIBentryALTinterwordspacing
M.~M. Wolf, ``Quantum channels \& operations: Guided tour,'' 2012. [Online].
  Available:
  \url{http://www-m5.ma.tum.de/foswiki/pub/M5/Allgemeines/MichaelWolf/QChannelLecture.pdf}
\BIBentrySTDinterwordspacing

\end{thebibliography}

\end{document}